\documentclass[lettersize,journal]{IEEEtran}
\usepackage{amsmath,amsfonts}
\usepackage{algorithmic}
\usepackage{algorithm}
\usepackage{array}
\usepackage[caption=false,font=normalsize,labelfont=sf,textfont=sf]{subfig}
\usepackage{textcomp}
\usepackage{stfloats}
\usepackage{url}
\usepackage{verbatim}
\usepackage{graphicx}
\usepackage{cite}
\usepackage{listings}
\usepackage{amsmath,amsfonts}
\usepackage{graphicx}
\usepackage{amsfonts}
\usepackage{array}
\usepackage{amssymb}
\usepackage{helvet}
\usepackage{color}
\usepackage{bm}
\usepackage{float}
\usepackage{cite}
\usepackage{multirow}
\usepackage{slashbox}
\usepackage{amssymb}
\usepackage[caption=false,font=normalsize,labelfont=sf,textfont=sf]{subfig}
\usepackage{amsmath,amsfonts}
\usepackage{algorithmic}
\usepackage{algorithm}
\usepackage{textcomp}
\usepackage{stfloats}
\usepackage{url}
\usepackage{verbatim}
\usepackage{graphicx}
\usepackage{cite}
\usepackage{amssymb}
\usepackage{makecell}
\usepackage{mathrsfs}

\newcommand{\RNum}[1]{\uppercase\expandafter{\romannumeral #1\relax}}

\begin{document}

\title{Integrated Sensing and Communications \\ Framework for 6G Networks}
\author{Hongliang Luo,  Tengyu Zhang, Chuanbin Zhao, Yucong Wang, Bo Lin,  \\ Yuhua Jiang,  Dongqi Luo, and Feifei Gao
        % <-this % stops a space
\thanks{H. Luo, T. Zhang, C. Zhao, Y. Wang, B. Lin,  Y. Jiang,  D. Luo, and F. Gao are with Department of Automation, Tsinghua University, Beijing 100084, China (email: luohl23@mails.tsinghua.edu.cn; 
zhang-ty22@mails.tsinghua.edu.cn;
zcb23@mails.tsinghua.edu.cn;
wangyuco21@ mails.tsinghua.edu.cn;
linb20@mails.tsinghua.edu.cn;
jiangyh20@mails.tsing hua.edu.cn;
dongqiluo@gmail.com;
feifeigao@ieee.org).}
% <-this % stops a space
%\thanks{Manuscript received April 19, 2021; revised August 16, 2021.}
}



\maketitle

\begin{abstract}

In this paper, we  propose a  novel integrated sensing and communications (ISAC) framework for the sixth generation (6G) mobile networks, 
 in which we decompose the real physical world into static environment, dynamic targets, and various object materials. 
The ubiquitous static environment occupies the vast majority of the physical world, for which we  design  
\emph{static environment reconstruction (SER)} scheme to 
obtain the  layout and point cloud information of  static buildings. 
The dynamic targets floating in static environments create the spatiotemporal transition of the physical world, for which we design  comprehensive \emph{dynamic target sensing (DTS)} scheme to  detect, 
estimate, track, image and recognize the dynamic targets in real-time. 
The  object materials enrich the electromagnetic laws of the physical world, for which 
we  develop 
\emph{object material recognition (OMR)} scheme to estimate the electromagnetic coefficient of the    objects.
Besides, to integrate these sensing functions into  existing communications systems, we  discuss the  interference issues and corresponding solutions for ISAC cellular networks. 
Furthermore, we  develop an ISAC hardware prototype platform that can reconstruct  the environmental maps and sense the  dynamic targets while maintaining communications  services. 
With all these designs, the proposed ISAC framework can support multifarious  emerging applications, such as 
 digital twins, low altitude economy, internet of vehicles,  marine management, deformation monitoring, etc. 

\end{abstract}

\section{Introduction}
As wireless communications and radar sensing  move towards a historic intersection,  
researchers are committed to integrating  communications  functions  and sensing  functions within the same  device in the sixth generation (6G)  wireless communications networks\cite{ISAC0202}. 
Generally, integrated sensing and communications (ISAC)  system aims  to multiplex the  time-space-frequency  resources and utilize communications  signals to sense the various information of  real physical world, such as building layout, target location, personnel activities, etc. 
%while  ensuring users (UEs) communications  performance.
Once  the  base stations (BSs) obtain the sensing  information, one can not only better serve communications users (UEs), but also support multifarious emerging  applications. 
With all these advantages, 
ISAC possesses sufficient potential to re-define the wireless networks, re-deploy the  air-interface technologies, and re-build the business ecosystem.
Therefore, ISAC has been officially approved by  ITU-R  IMT-2030 as one of  six decisive usage scenarios for  6G networks.

Currently, numerous  signal processing techniques have been designed  for ISAC systems to support  abundant sensing functionalities.  
For example, the authors of  \cite{10271123}  propose a  low-overhead UE localization method  with orthogonal frequency division multiplexing (OFDM) signals.  
The authors of \cite{10048770}  provide a high-precision target parameter estimation technology based on   multiple signal classification (MUSIC) algorithm.
In \cite{9947033}, the authors design   dynamic targets tracking scheme with  extended Kalman filter for vehicle networking scenarios.
In \cite{9727176}, the authors  construct the environmental maps using OFDM signals, which is commonly referred to as
simultaneous localization and mapping (SLAM).
In   \cite{10333765}, the authors  consider the joint use of  BSs and unmanned aerial vehicles (UAVs) to realize object imaging. 
More interestingly, \cite{afzal2023agritera} 
senses the ripeness of fruits with 
 terahertz frequency band. 
However, these studies mainly focus on  single sensing task like target detection, or parameter estimation, or tracking, while they are dispersed and lack a coherent overarching theme. With the commercialization of 5G-Advanced and the acceleration of 6G research, it is essential for academic and industrial communities to  sort out of the mainline to classify these  works.

\begin{figure*}[!t]
	\centering
	\includegraphics[width=170mm]{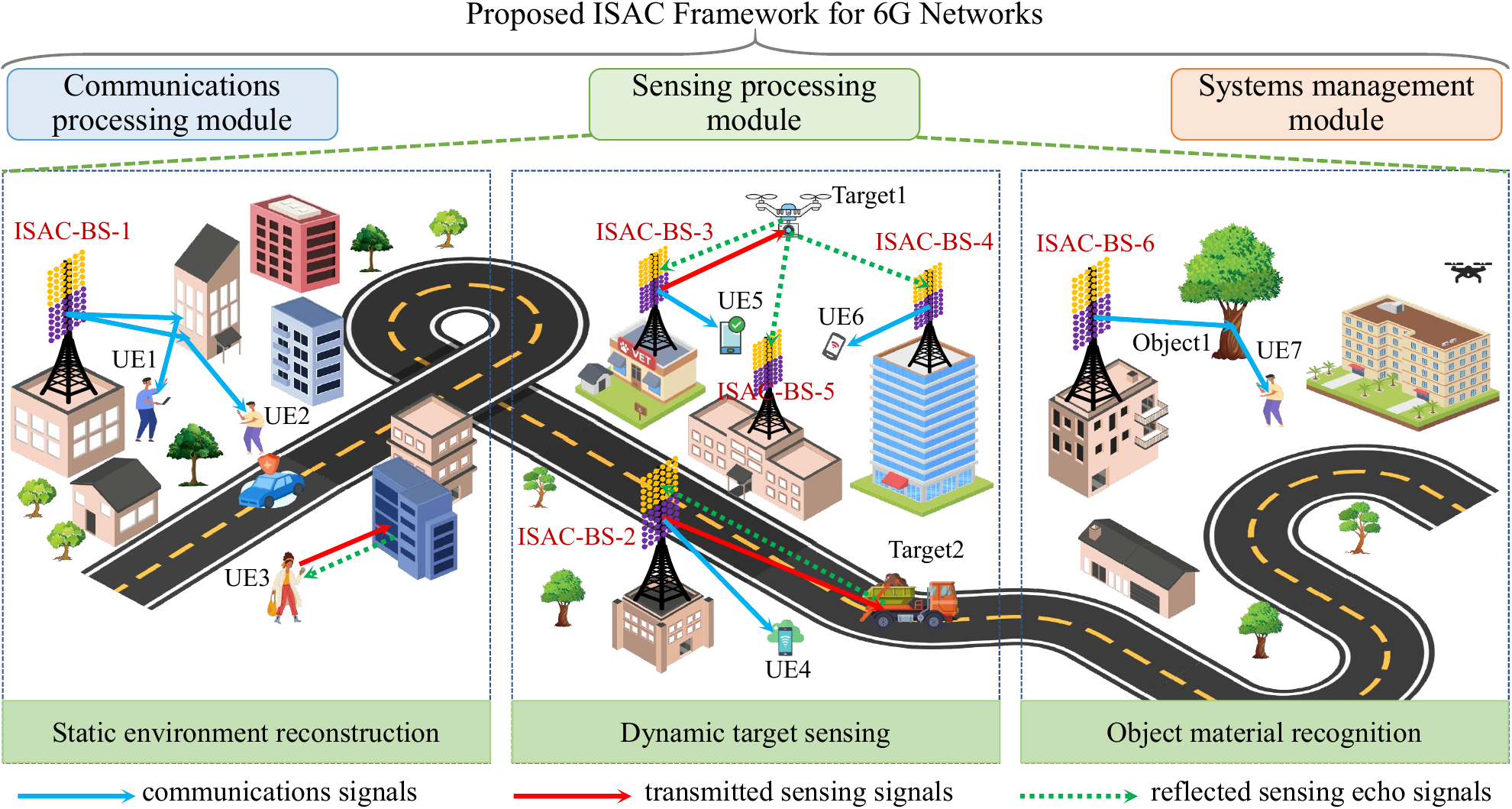}
	\caption{The  architecture of  the proposed ISAC  framework for 6G networks.}
	\label{fig_1}
\end{figure*}

Researchers are also  developing the  resource allocation, waveform design, and interference management schemes to integrate the aforementioned sensing functions into the existing communications systems. A popular solution is to utilize orthogonal time-space-frequency resources for communications and sensing separately\cite{ISAC0201}. 
In time-division ISAC mode,  BSs periodically deploy the communications function and sensing functions on different time blocks, which has been  widely adopted by the industry. 
In frequency-division ISAC mode,  BSs use  different subcarriers of OFDM signals for  communications and sensing. 
In space-division ISAC mode, BSs  simultaneously generate  communications beams and sensing beams, which is well suited  for massive multi-input and multi-output (MIMO) systems that inherently possess  abundant space resources.  
However, these schemes are only  well  studied on the single BS. 
With  further increase of BSs density in 6G and the addition of  collaborative sensing capabilities, ISAC  networks are facing new interference problems.

In this paper, we  propose a  novel ISAC framework for 6G  networks, in which we decompose the sensing of the  real physical world into separately  sensing the 
static environment, dynamic targets, and  object materials. 
For static environment reconstruction, we  obtain the location and point cloud information of the  buildings with multi-UEs and  multi-BSs cooperation. 
For dynamic target sensing, we develop the real-time  comprehensive sensing schemes, including many key components like clutter suppression, target detection, parameter estimation, track  management, and target imaging, etc.
For object material recognition,  we design the  electromagnetic constant estimation method based on compressive sensing, and then  utilize the estimation results to identify the  materials of  object.
Besides, we  discuss the  interference issues and corresponding solutions for ISAC cellular networks. 
Furthermore, we  develop an ISAC hardware prototype platform that can reconstruct   environmental maps and sense   dynamic targets  while maintaining UEs communications. 
The proposed ISAC framework can support various emerging applications, such as 
 digital twins, low altitude economy, internet of vehicles, smart factory, marine management, deformation monitoring, etc.

\section{Proposed ISAC  Framework}

Let us consider a typical 6G  scenario as shown in Fig.~1, where multiple  ISAC BSs and a large number of UEs are deployed to realize communications and sensing functions. Without loss of generality,
 we assume that the system is working under OFDM modulation.  
To match the massive number of array antennas in 6G networks,   
 we here mainly  focus on  the space-division ISAC systems, and 
 assume that each BS consists of one {hybrid unit (HU)} and one {radar  unit (RU)}.  HU is responsible for transmitting downlink communications signals and receiving uplink communications signals, as well as transmitting downlink sensing signals;  RU is responsible for receiving the sensing echo signals.

%HU simultaneously transmits the communications beams and sensing beams, while RU  generates the sensing beams to receive echo signals.

As described in Fig.~1, the proposed ISAC framework consists of three key modules.  
\emph{Communications processing module (CPM)} can complete the uplink and downlink communications tasks between BSs and UEs, which has been well studied in previous networks and will be omitted in this paper. 
\emph{Sensing processing module (SPM)} with multiple basic sensing  functions aims to  obtain   various information of  real physical world, which will be introduced in Section~\RNum{3}.
\emph{Systems management module (SMM)} targets at  effectively integrating SPM into existing communications  networks by developing  resource scheduling, waveform design, and interference management schemes,  which will be introduced in Section~\RNum{4}.
Based on CPM, SPM, and SMM, one may  reconstruct the real physical world with various sensing results, and further enhance systems management and communications transmission.

\section{Sensing Processing Module}

In physical world,  the reflection and scattering laws caused by  environment and moving targets  determine the direction of electromagnetic wave propagation, while
the electromagnetic coefficients of objects with different materials  further determine the amplitude and phase of electromagnetic waves after reflection and scattering.
Thus the physical world has three major components with different characteristics:
1) static environments, such as roads,  buildings and trees; 
2) dynamic targets, such as pedestrians, vehicles and  UAVs; 
3) various materials possessed by the environments or targets, such as steel material, wood material and fabric material.  
Correspondingly, we  divide  SPM into three basic sensing  functions, including  \emph{static environment reconstruction (SER)}, \emph{dynamic target sensing (DTS)}, and \emph{object material recognition (OMR)}.
The static environments possess gradual variability and SER mainly relies on non-line-of-sight (NLoS) channel decomposition; dynamic targets possess temporal variability and DTS mainly relies on line-of-sight (LoS) channel decomposition;  OMR mainly relies on electromagnetic calculations. 
We will next  develop different sensing schemes to realize SER, DTS, and OMR, respectively.

\subsection{Static Environment Reconstruction}

\begin{figure*}[!t]
	\centering
	\includegraphics[width=182mm]{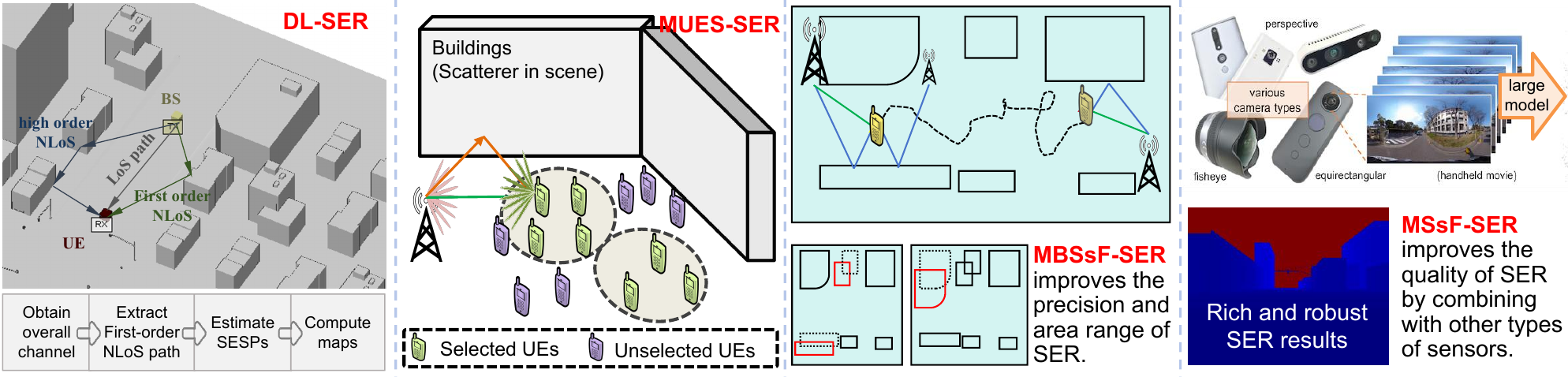}
	\caption{Schematic diagram of static environment reconstruction.}
	\label{fig_1}
\end{figure*}

Since ubiquitous static environments occupy the vast majority of the physical world,  SER aims to sense the two-dimensional (2D) or three-dimensional (3D) point cloud information of static environments, mainly including the mapping of building layouts.
SER can assist ISAC systems in supporting various vertical applications related to surveying and mapping,  driving navigation,  etc. 
Since the changes of  static environment are usually slow, SER can be updated at relatively long intervals,
and the sensing  results can be reused.
For example, ISAC systems can perform SER once a month or once a week. Due to the insensitivity of  radio wave to lighting conditions, the systems could  update the results of SER in  late night to avoid peak resource usage.

Generally, the communications channels between BSs and UEs comprise the layout information of  static environments. Especially, the  precise positions and other information of the static environment scattering points (SESPs)   can be elaborately obtained from
the first-order NLoS paths between BSs and UEs.
Once  ISAC systems obtain a large number of SESPs, it  can reconstruct high-precision maps.
Fig.~2 shows four potential technologies for  SER, which are  introduced as follows.

{\bf Deep learning based SER   (DL-SER)}\cite{2023arXiv230302617M}.
DL-SER provides the basic idea of mapping with  single BS and single UE.  
At a certain moment,  UE is considered stationary,   and then one  can extract the sub-channels corresponding to the first-order NLoS paths from the overall channel between BS and UE with  DL algorithms. 
Then   low rank decomposition and channel inversion are performed  to obtain the positions  of  SESPs.
As UE moves, the number of SESPs gradually increases, and then 
the  mapping  results of  static environment are obtained through   global geometric calculations.

{\bf Multi-UEs selection based SER (MUES-SER)}\cite{2024arXiv240317810L}.
Note that DL-SER with single UE  critically relies on UE's movement to form multitudinous SESPs, 
 and the uncontrolled movement trajectory   will  also restrict the area range of mapping results.
Fortunately,  in  multiple UEs scenario, 
MUES-SER can create one or minor  \emph{virtual UE} trajectories by selecting the most suitable UEs subset  from multitudinous UEs, 
and   implement   DL-SER  on each virtual UE to obtain the local mapping results of single virtual UE. 
Next, the multi-UEs fusion algorithm based on evidence theory can be designed to obtain the  large area range and high precision mapping results.

{\bf Multi-BSs fusion based SER (MBSsF-SER).} In the scenario of multi-BSs collaborative sensing, the local  mapping results obtained by each single BS through DL-SER or MUES-SER will be synchronized to the data fusion center. Then the  data fusion center performs noise  suppression, ghost point elimination, and power measurement alignment to preprocess these local maps, and futher obtains the multi-BSs fusion mapping results based on  point cloud feature matching and situation computing within local maps.

{\bf Multi-sensors fusion based SER (MSsF-SER).}
MSsF-SER can  significantly improve the quality of mapping results by combining  different advantages of ISAC systems and other types of sensors, such as cameras, LiDAR, etc. For example,  MSsF-SER that combines ISAC systems and cameras leads to a threefold accuracy improvement over the visual only method\cite{2024arXiv240317810L}, thus increasing the accuracy and robustness of environment reconstruction, especially under challenging weather conditions like raining and snowing.

\begin{figure*}[!t]
	\centering
	\includegraphics[width=180mm]{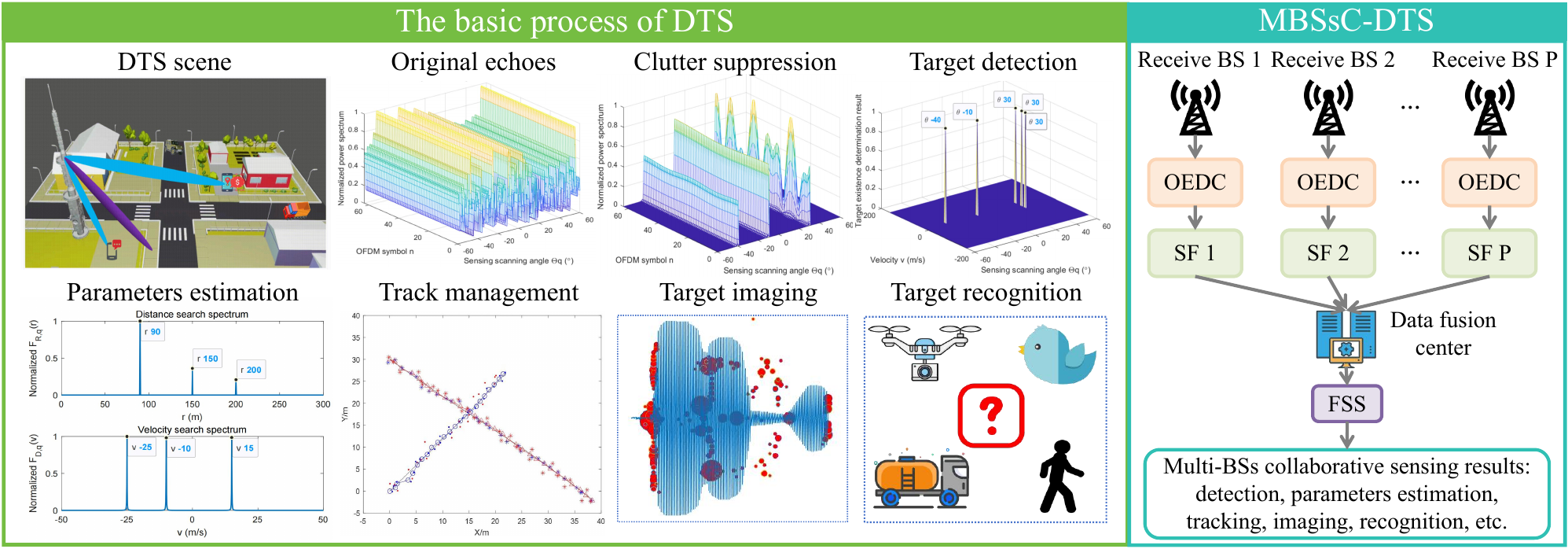}
	\caption{Schematic diagram of dynamic target sensing.}
	\label{fig_1}
\end{figure*}

\subsection{Dynamic Target Sensing}

The dynamic targets floating in static environments create the spatiotemporal transition of  physical world. DTS
includes the discovery, detection, parameter estimation,   tracking,  imaging, recognition,   of the  targets, 
 which  helps ISAC systems support various scenarios related to targets motion, such as low-altitude economy, vehicle networking,  smart factories,  etc. 
Different from  static environments, the changes of dynamic targets are usually rapid, and thus it is necessary to develop  specialized and real-time DTS algorithms.

We propose to divide the  process of DTS into three stages. 
1) BS first  enables \emph{search stage}, during which  it  employs  sensing beam  searching the whole  area 
at certain angle intervals to  discover and detect   dynamic targets, and preliminarily estimate the motion parameters of  targets;
2) BS enters \emph{tracking stage}, during which it predicts the motion trajectories of targets with   methods like Kalman filtering, and generates one or more sensing beams to track these targets;
3)  BS can even perform  \emph{imaging and recognition  stage}, during which it mobilizes more resources to image  targets of interest, or identify the category of target, like  vehicle, UAV, etc.

To support these sensing stages, Fig.~3 shows the basic process of DTS, which are introduced  as follows.

{\bf Clutter suppression.}  The  echoes received by  BS not only includes  dynamic target echoes of interest, but also plentiful  environmental clutter caused by  static environments.  Since the  clutter will cause serious negative interference to DTS, it is necessary to adopt  clutter suppression algorithms to suppress clutter from  original echoes   and extract  dynamic target echoes.
The main principle   is to utilize the difference in Doppler frequency shift between clutter and dynamic target echoes to filter out the clutter in OFDM symbol  or Doppler domain\cite{10477890}.

{\bf Target detection.}   Target detection aims  to use the  characteristics of effective signals and noise  to determine  whether the target  is present or not.
ISAC BS can utilize the multi-subcarrier mechanism of OFDM signals to achieve better target detection performance than traditional radar system by combining individual detection results on all subcarriers\cite{10477890}.

{\bf Parameter estimation.} Parameter estimation refers to the estimation of  position,  velocity, and radar cross section (RCS)  of the target.  MIMO-OFDM-ISAC system mainly relies on multiple antennas  to estimate the target's angle, relies on multiple subcarriers to estimate the target's distance, relies on multiple OFDM  symbols to  estimate the target's velocity, and relies on  channel power changes to estimate the target's RCS. Currently, various  estimation methods based on FFT, MUSIC, ESPRIT, tensors, have been derived for ISAC system.
More excitingly,  with the help of large-scale MIMO, \cite{2023arXiv231216441L}  expands the traditional 4D MIMO radar that can only estimate the horizontal angle, pitch angle, distance, and radial velocity of target to an innovative 6D MIMO radar that can additionally  estimate the horizontal angular velocity and pitch angular velocity of target.

{\bf Track management.}  Track management aims to identify  the  trajectory of target from the  points detected by BS, as well as  tracking and predicting    target's motion, which mainly includes track initiation, track  association, and trajectory tracking.
 Generally, BS needs to use  Kalman filtering  or DL   based methods  to track and predict the trajectory and  parameters changes of target, and then adjusts the direction or even the beam width of  the 
 tracking beam  in real-time.

{\bf Target imaging.} Target imaging is also known as target reconstruction. Synthetic aperture radar (SAR)  and inverse synthetic aperture radar (ISAR) technology can be utilized in   ISAC system to image the targets, i.e., obtaining the point cloud information of extended targets and further analyzing the topological structure  of target. 

{\bf Target recognition.} Target recognition refers to identifying the categories of important targets, such as pedestrians, vehicles, UAVs, birds, etc. The ISAC system can accurately identify the targets by utilizing the differences in the features such as imaging results, RCSs, micro Doppler spectrums, high resolution range profiles (HRRPs),  motion trajectories, etc.

On the basis of DTS with single BS, DTS with multi-BSs collaboration (MBSsC-DTS) demonstrates lots of advantages like improved sensing accuracy and range, as well as continuous tracking across regions. 
Specifically,  ISAC networks first determine the transmission and reception nodes, and  perform signals time synchronization and beam space synchronization among multi-BSs.
Next, each single BS  employs  appropriate sensing   processing strategy to achieve preliminary  processing of  original echoes, which is  termed as \emph{data compression of  original echoes (OEDC)}. Then each single BS transmits the \emph{sensing  features (SF)} obtained from OEDC to the data fusion center, where the  suitable fusion sensing  scheme (FSS) is  adopted  to realize MBSsC-DTS. 
Note that, the three key steps in MBSsC-DTS:  the design of OEDC, the selection of SF, and the devise of FSS, usually constrain each other and thus require sophisticated  consideration.

\begin{figure*}[!t]
	\centering
	\includegraphics[width=182mm]{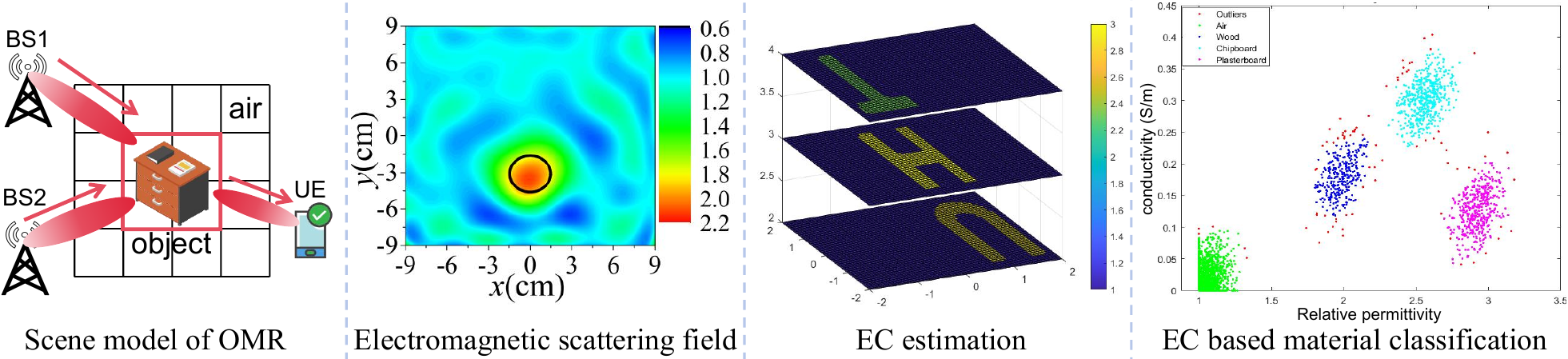}
	\caption{Schematic diagram of object material recognition.}
	\label{fig_1}
\end{figure*}

\subsection{Object Material Recognition}

The objects of different materials in either  static environment or  dynamic target possess different electromagnetic coefficients, whose absorption, reflection, and scattering of electromagnetic signals are different. It is believed that various object materials enrich the electromagnetic laws of  physical world. 
Unfortunately,  existing  ISAC works overlooked the sensing of object's electromagnetic coefficients.
To fill  this research gap, we propose OMR to identify the material of  object by  estimating its electromagnetic coefficient, which will  provide a new paradigm  for ISAC  to  interpret the world, such as helping communications systems rebuild more accurate twin electromagnetic channels, 
identifying the categories of objects, discovering dangerous utensils, and even judging the ripeness of fruits. 
Fig.~4 shows the basic model and key components of OMR, which are introduced as follows\cite{10538217,2024arXiv240506364J}.

{\bf Electromagnetic inverse scattering model of OMR.}  The electromagnetic coefficient of  object is characterized by \emph{electromagnetic contrast (EC)}, which is composed of both relative permittivity and conductivity.
According to the laws of electromagnetics, 
one can use  Lippmann-Schwinger equation to describe the electromagnetic field  channel  containing the EC of scatterers from BS to  scatterer and then to UE.
The  electromagnetic inverse scattering model of  OFDM system can be obtained by modeling the electromagnetic scattering channel  of each  subcarrier.

{\bf Compressive sensing based  EC estimation.}  Since the EC of  object has sparsity in  sensing domain,   compressive sensing (CS) methods can be used to estimate the EC parameter.
Specifically, after separating the imaginary and real parts of the electromagnetic inverse scattering model,  EC estimation  can be transformed into an equivalent mixed norm minimization problem, which can be solved by  spectral projection gradient method. 
Then one can obtain the relative permittivity and conductivity of object based on the estimated EC.

{\bf  EC based material classification.} 
After obtaining the EC of the  objects, one  can use the   relative permittivity and conductivity  as the data features,  and employ the K-means algorithm to cluster a large number of objects as shown in Fig.~4. 
In addition, support vector machines (SVM), convolutional neural networks (CNN), and other AI methods can be used to realize  material classification and recognition of objects  based on the data characteristics of electromagnetic coefficients.

\section{Systems Management Module}

Unlike traditional radar systems, ISAC systems could utilize densely deployed BSs networks for collaborative sensing, thereby improving the  sensing accuracy and coverage range. However, there would exist  not only communications interference and sensing interference among multiple BSs, but also mutual interference between communications and sensing among multiple BSs\cite{2024arXiv240316189N}.
To address these interference issues, SMM should fully utilize the  environmental information and data sharing between BSs, and comprehensively schedule  time-space-frequency resources to ensure the effective integration of the  SER, DTS, and OMR functions into the existing communications  network.
Some discussions about SMM are provided  as follows.

{\bf Multi-BSs deployment and on-off control.}
The 6G networks will re-deploy a large number of ISAC BSs, whose locations and on-off strategies  will directly affect the effectiveness  of interference suppression in ISAC cellular networks. Specifically, after obtaining the environmental map of  service area, one  can use the heuristic algorithms to obtain the optimal locations for deploying BSs by minimize the numerical interference  between communications and sensing among multiple BSs. Besides, by modeling the on-off problem of multiple BSs as the  reinforcement learning problem, it is possible to optimize the real-time communications and sensing performance  of   networks while saving the  energy consumption of the systems.

{\bf  Collaborative beamforming.}
After determining the BSs' locations,  ISAC systems need  to simultaneously generate both communications beams and   sensing beams. ISAC networks should avoid the serious interference between sensing beams from multiple BSs by designing collaborative sensing scanning strategies and active avoidance mechanisms. 
Besides,  ISAC networks can jointly optimize the precoding matrix for communications beams and sensing beams among multiple BSs, thereby maximizing  ISAC performance indicators, including signal-to-noise-ratio (SNR), equivalent sensing power, etc.

{\bf  Joint power and bandwidth allocation.}  Although joint precoding  optimization can effectively reduce  networks interference, the accompanied  high computational complexity makes it difficult to implement  in  complex scenarios. As a compromise between performance and complexity,
after modeling power allocation as a continuous variable and bandwidth allocation as a binary discrete variable, one can reduce the networks interference and improve ISAC performance by alternately optimizing the beams power and bandwidth resource allocation among multiple BSs. 

\begin{figure*}[!t]
	\centering
	\includegraphics[width=180mm]{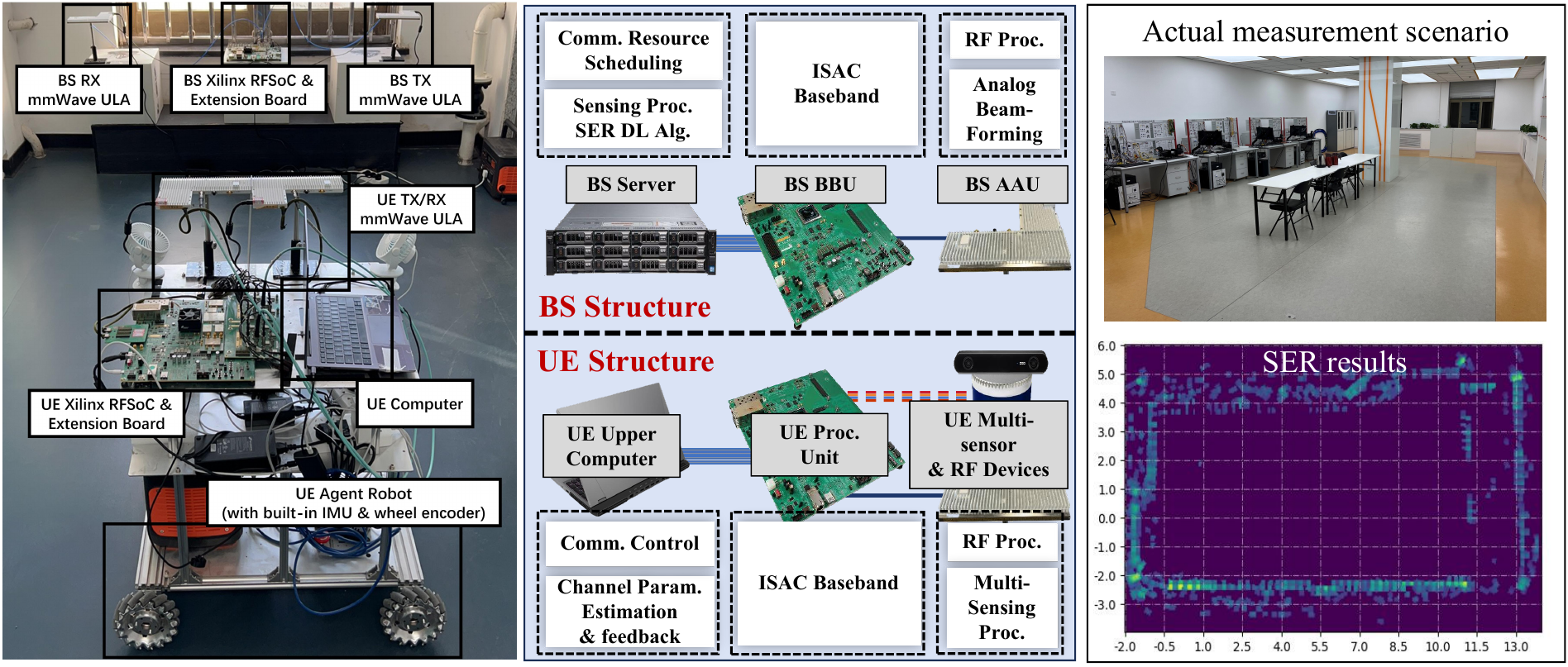}
	\caption{Hardware prototype  of static environment reconstruction.}
	\label{fig_1}
\end{figure*}

\begin{figure*}[!t]
	\centering
	\includegraphics[width=180mm]{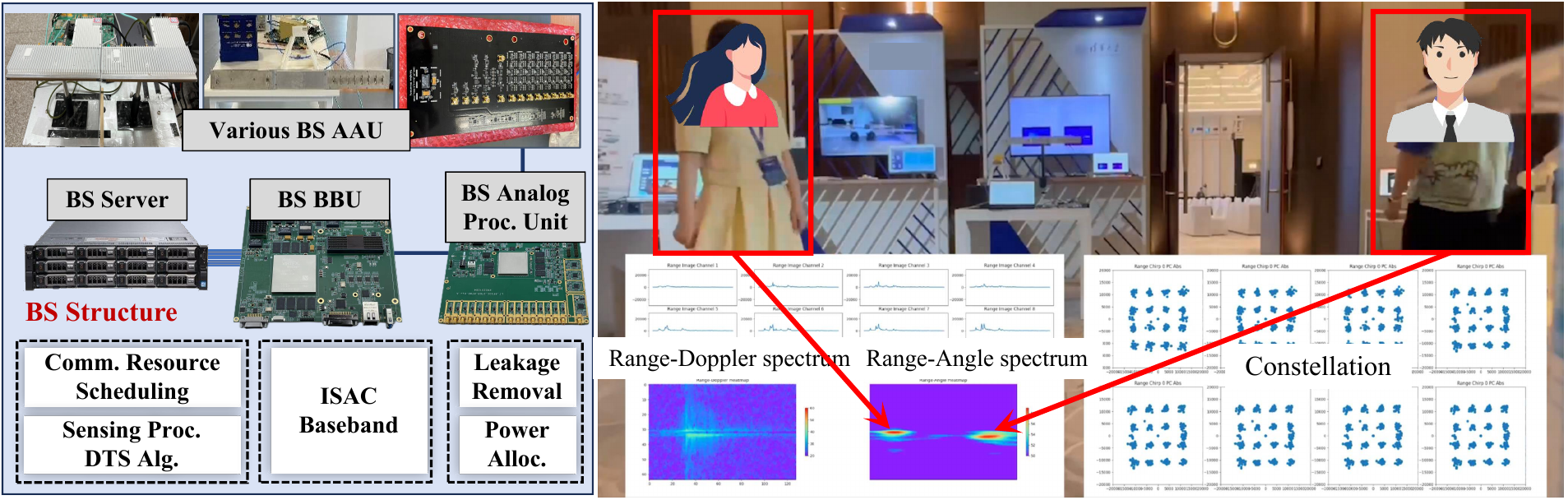}
	\caption{Hardware prototype  of dynamic target sensing.}
	\label{fig_1}
\end{figure*}

\section{ISAC Hardware Prototype}

A carefully designed hardware platform is the important guarantee for the smooth commercial deployment of ISAC. 
We   develop the ISAC hardware prototype based on field-programmable gate array (FPGA), which can support  large bandwidth and flexible secondary design.
Here we present the hardware prototypes for SER and DTS,  respectively.

The  hardware prototype for  SER is shown in Fig.~5. 
Since the core of SER lies in precisely extracting and estimating the channel parameters for NLoS paths through communications between BS and UE,  we  set up the BS and UE   hardware prototypes working in   $26$~GHz frequency band, with  working bandwidth of $820$~MHz.
Specifically, we deploy the communications resource scheduling scheme and DL-SER algorithm  on the server of BS,  develop  ISAC baseband algorithms for the baseband unit (BBU) of BS, and build the radio frequency (RF) front-end for active antenna unit (AAU) of BS  through an  uniform linear array (ULA) with $32$ antennas.  
On the other side, we deploy the communications control, channel parameter estimation, and feedback schemes on the upper computer unit of UE, design the ISAC baseband algorithms for the processing unit of UE, and build the RF front-end of UE through an ULA with $32$ antennas. 
Besides, we  equip the UE with inertia measurement unit (IMU) and wheel encoder (WE) to improve UE localization accuracy.
Fig.~5 shows the SER results  with the simultaneous   communications between BS and UE.
The communications rate of the system is maintained above $1.5$ Gbps,
the measured sensing angle resolution  and distance resolution are $3^\circ$ and $0.183m$, respectively. The average error of UE  localization is $0.0347 m$, and the average error of SER results is $0.2610 m$. These results indicate that the designed  prototype  can accurately construct the environmental map while maintaining UE communications.
Besides, the precision of SER can be further improved with the increase of   system bandwidth and the number of array antennas.

The hardware prototype  for  DTS is shown in Fig.~6, where  the BS  is working in   $5.5$~GHz frequency band, with   bandwidth  $820$~MHz.
We deploy the communications resource scheduling scheme and DTS  algorithm  on the server of BS,  develop  ISAC baseband algorithms for the BBU of BS, and build the RF front-end for AAU of BS  through a receiving-ULA with $8$ antennas and a transmitting-ULA with $2$ antennas.  
Besides, $64$ OFDM symbols are used for DTS. 
Fig.~6 simultaneously shows the constellation diagram of BS communications, as well as the range-Doppler spectrum and range-angle spectrum of DTS. 
The measured sensing angle resolution, distance resolution, and velocity resolution are $3^\circ$, $0.183m$ and $0.42m/s$, respectively. 
These results indicate that the designed  prototype  can effectively sense the dynamic targets,  while maintaining the  communications between BS and UE.

\section{Conclusions}

In this paper, we have  proposed a  novel ISAC framework for  6G mobile networks, by decomposing  the sensing of the  real physical world into separately  sensing the 
static environment, dynamic targets, and various object materials. 
For SER,  we have designed  DL based, multi-UEs selection based, multi-BSs fusion based, and multi-sensors fusion based SER technologies to obtain the location  and point cloud information of the buildings. 
For DTS, we have developed the comprehensive sensing schemes based on OFDM signals for single BS and multi-BSs to 
detect, locate, track, image, and recognize the dynamic targets in real-time. 
For OMR,  we have  designed the electromagnetic constant estimation method based on compressive sensing, and further utilized the estimation results to identify the materials of object. 
To integrate the above sensing functions into existing communications  systems, we have discussed the interference issues and corresponding solutions for ISAC networks. 
Furthermore, we have  developed an ISAC hardware prototype platform that could reconstruct   environmental maps and sense   dynamic targets  while maintaining  communications. 
With all these designs, the proposed  framework can support
ISAC system in  constructing the mapping relationship from  real physical world to  digital twin world while ensuring communications, thereby empowering various emerging applications, such as channel twins, low altitude economy, internet of vehicles, etc.

\bibliographystyle{ieeetr}
\bibliography{paper10.bib}

\vfill

\end{document}